\documentclass[12pt]{article}

\usepackage{epsf}
\usepackage{amsmath2000}
\allowdisplaybreaks

\begin{document}

\baselineskip=17.5pt plus 0.2pt minus 0.1pt

\makeatletter
\@addtoreset{equation}{section}
\renewcommand{\theequation}{\thesection.\arabic{equation}}
\renewcommand{\thefootnote}{\fnsymbol{footnote}}
\newcommand{\bra}[1]{\langle #1 |}
\newcommand{\ket}[1]{| #1 \rangle}
\newcommand{\bracket}[2]{\langle #1| #2 \rangle}
\newcommand{\lvac}{\langle 0 |}
\newcommand{\rvac}{| 0 \rangle}
\newcommand{\num}[1]{{}_{#1}}
\newcommand{\del}{\partial}
\newcommand{\Le}{\left}
\newcommand{\Ri}{\right}
\newcommand{\nn}{\nonumber}
\newcommand{\half}{\frac{1}{2}}
\newcommand{\ve}[1]{\boldsymbol{#1}}
\newcommand{\refb}[1]{(\ref{#1})}
\newcommand{\p}{\partial}
\newcommand{\cI}{\mathcal{I}}
\newcommand{\cV}{\mathcal{V}}
\newcommand{\cN}{\mathcal{N}}
\newcommand{\Nc}{\mathcal{N}_\mathrm{c}}
\newcommand{\cQ}{\mathcal{Q}}
\newcommand{\cM}{\mathcal{M}}
\newcommand{\Psic}{\Psi_\mathrm{c}}
\newcommand{\Psicm}{\Psi_\mathrm{c}^\mathrm{m}}
\newcommand{\Psicg}{\Psi_\mathrm{c}^\mathrm{g}}
\newcommand{\Phim}{\Phi^\mathrm{m}}
\newcommand{\Phit}{\Phi_\mathrm{t}}
\newcommand{\Phik}{\Phi^{(k)}}
\newcommand{\rhop}{(\rho_+)}
\newcommand{\rhom}{(\rho_-)}
\newcommand{\fz}{\ve{f}^{(0)}}
\newcommand{\bkl}{\beta^{\mu\nu}(\kappa,\lambda)}
\newcommand{\gkl}{\gamma^{\mu\nu}(\kappa,\lambda)}
\newcommand{\wtcQ}{\cQ_{\rm B}}
\newcommand{\Vm}{V^{\mathrm{m}}}
\newcommand{\sPmatrix}[1]{
            \left(\begin{smallmatrix} #1 \end{smallmatrix}\right)}
\newcommand{\e}{e}


\begin{titlepage}
\title{
\hfill\parbox{4cm}
{\normalsize KUNS-1797\\{\tt hep-th/0208067}}\\
\vspace{1cm}
{\bf Higher Level Open String States\\ 
from Vacuum String Field Theory
}
}
\author{
Hiroyuki {\sc Hata}\thanks{
{\tt hata@gauge.scphys.kyoto-u.ac.jp}},
\
and
\
Hisashi {\sc Kogetsu}\thanks{
{\tt kogetsu@gauge.scphys.kyoto-u.ac.jp}}
\\[15pt]
{\it Department of Physics, Kyoto University, Kyoto 606-8502, Japan}
}
\date{\normalsize August, 2002}
\maketitle
\thispagestyle{empty}

\begin{abstract}
\normalsize
We construct massive open string states around a classical
solution in the oscillator formulation of Vacuum String Field Theory.
In order for the correct mass spectrum to be reproduced,
the projection operators onto the modes of the left- and right-half of
the string must have an anomalous eigenvalue $1/2$, and
the massive states are constructed using the corresponding
eigenvector.
We analyze numerically the projection operators by regularizing them
to finite size matrices and confirm that they indeed have eigenvalue
$1/2$. Beside the desired massive states, we have spurious massive as
well as massless states, which are infinitely degenerate.
We show that these unwanted states can be gauged away.
\end{abstract}
\end{titlepage}

\section{Introduction and summary}

Vacuum String Field Theory (VSFT)
\cite{RSZ:0012251,RSZ:0102112,Rastelli:2001vb,RSZ:0106010}
has been proposed as a string field
theory expanded around the tachyon vacuum.
In order for VSFT to really be connected to ordinary bosonic string
theory on an unstable D25-brane,
there must exist a Lorentz and translationally invariant
classical solution of VSFT satisfying the following two requirements:
the fluctuation modes around the solution reproduce the open
string spectrum, and the energy density of the solution is equal
to the D25-brane tension.
Recently, there has been much progress in understanding the above
problem.
In particular, a full classical solution of VSFT including the ghost
part has been presented in \cite{HK:0108150} using the oscillator
formulation, and 
the tachyon and the massless vector fluctuation modes have been
constructed there.
They have shown that the tachyon mass is correctly
reproduced. However, the massless vector mode contains an arbitrary
vector in the level number space, implying that there are infinite
number of massless vector states. This problem was later resolved by
Imamura \cite{Imamura:0204031}: most of the massless vector modes can
be gauged away by VSFT gauge transformation leaving only one physical
vector mode.

The purpose of this paper is to carry out the oscillator construction
of fluctuation modes representing higher level massive modes of open
string.\footnote{
In this paper we shall consider only the oscillator formulation of
VSFT.
For the construction of the fluctuation modes using boundary
conformal field theory, see
\cite{Rastelli:2001wk,Rashkov:2001js,Rashkov:2002xz,Okawa:0204012}.
}
This is in fact a non-trivial and interesting problem.
Let us take, as a candidate massive state with mass
squared equal to $(k-1)\alpha'$, a state given as $k$ matter
creation operators $a_n^{\mu\dagger}$ acting on the tachyon state.
It is a natural extension of the tachyon and massless vector states
of \cite{HK:0108150}. 
However, naive analysis shows that this kind of states are all
massless.
The wave equation (namely, the linearized equation of motion),
$\wtcQ\Phi=0$, for the fluctuation 
$\Phi$ of the above type is reduced to a simple algebraic equation
consisting only of the projection operators, $\rho_+$ and $\rho_-$,
onto the modes of the left- and right-half of the string
\cite{Rastelli:2001rj}.
The masslessness is a consequence of the basic property of projection
operators, $\rho_\pm^2 =\rho_\pm$.

We find, however, that the above mode can represent a massive state
with the expected mass squared, $(k-1)\alpha'$, if $\rho_\pm$ has an
anomalous eigenvalue $1/2$ despite that it is a projection operator.
Such an anomalous eigenvalue is of course impossible for projection
operators in a finite dimensional space.
However, there is a subtle point for $\rho_\pm$ which is an operator
in the infinite dimensional space of string level number.
In fact, the eigenvector $\ve{f}^{(\kappa)}$ of the matrix
representation of the Virasoro algebra $K_1=L_1+ L_{-1}$ (the
eigenvalue $\kappa$ is continuous and extending from $-\infty$ to
$\infty$) is at the same time the eigenvector of $\rho_\pm$ and the
corresponding eigenvalue is the step function $\theta(\pm\kappa)$
\cite{RSZ:0111281}.
Therefore, the vector $\fz$ is the eigenvector of $\rho_\pm$ with
eigenvalue $\theta(0)$, which is indefinite but could be the desired
value $1/2$.
To verify whether this expectation is correct, we have to study
$\rho_\pm$ with some kind of regularization.
In this paper, we analyze numerically the eigenvalue problem of
$\rho_\pm$ regularized to finite size matrices to obtain results
supporting the above expectation: $\rho_\pm$ has an eigenvalue and
an eigenvector which tend to $1/2$ and $\fz$, respectively, as the
size of the matrices is increased.

Even if $\rho_\pm$ has the expected anomalous eigenvalue $1/2$, there
still remains a problem to be solved for the construction of higher
level open string modes. Analysis of the wave equation for
fluctuations of the above type, $(a^\dagger)^k\ket{\rm tachyon}$,
shows that there still exists infinite degeneracy of
massive states with mass squared equal to $\ell\alpha'$
($\ell\le k-2$).
In addition, we also have spurious massless states
mentioned above. We have to show that these unwanted states are not
physical ones. This problem is solved in the same manner as in the
massless vector case \cite{Imamura:0204031}:
infinite number of spurious states can be gauged away. However, we
need gauge transformations of a different kind from that used in
\cite{Imamura:0204031} in order to remove all the unwanted massive
states.

Our construction of massive open string states is not complete,
and there remain a number of future problems.
First, we have to present a rigorous analytic proof of the existence
of the anomalous eigenvalue $1/2$ of the projection operators
$\rho_\pm$. Second, as we shall see later, we construct only the
highest spin states at a given mass level. The construction of lower
spin states is our remaining subject.
Finally, in our analysis we consider the wave equation $\wtcQ\Phi=0$
only in the Fock space of first quantized string states. Namely, our
massive modes $\Phi$ satisfy the wave equation in the
sense of $\bra{\rm Fock} \wtcQ \ket{\Phi} = 0$ for any Fock space
element $\bra{\rm Fock}$. However, analysis of the potential height
problem of the D25-brane solution of VSFT shows that we have to
consider the equation of motion in a larger space including the states
constructed upon the D25-brane solution
\cite{Rastelli:2001wk,Okawa:0204012,Reexam}.
This is our important future problem.

The organization of the rest of this paper is as follows.
In sec.\ 2, we analyze the equation of motion for our candidate
massive modes and argue that $\rho_\pm$ needs an anomalous eigenvalue
$1/2$.
In sec.\ 3, we present numerical analysis of the eigenvalue problem of
finite size $\rho_\pm$.
In sec.\ 4, we show that the spurious states are unphysical ones
which can be removed by gauge transformations.
In appendix \ref{evaluation}, we present technical details used in the
text.

\section{Massive modes}\label{section-massive}

The action of VSFT is given by
\cite{RSZ:0012251,RSZ:0102112,RSZ:0106010}
\begin{equation}
 S=-K\Le(\half\Psi\cdot\cQ\,\Psi + \frac13\,\Psi\cdot(\Psi*\Psi)\Ri),
\label{eq:VSFT-action}
\end{equation}
where $K$ is a constant and the BRST operator $\cQ$ of VSFT consists
purely of ghost oscillators:
\begin{equation}
\cQ=c_0+\sum_{n=1}^\infty f_n\left(c_n +(-1)^n c_n^\dagger\right) .
\label{eq:cQ}
\end{equation}
The VSFT action (\ref{eq:VSFT-action}) is invariant under the gauge
transformation,
\begin{equation}
\delta_\Lambda\Psi=\cQ\Lambda+\Psi*\Lambda-\Lambda*\Psi.
\label{eq:dLPsi}
\end{equation}
The D25-brane configuration of VSFT is a translationally and Lorentz
invariant solution $\Psic$ to the equation of motion:
\begin{equation}
\cQ\Psic +\Psic *\Psic=0 .
\label{eq:EOM}
\end{equation}
Assuming that $\Psic$ factorizes into the matter part $\Psicm$ and the 
ghost one $\Psicg$, $\Psic=\Psicm\otimes\Psicg$,
(\ref{eq:EOM}) is reduced into the following two:
\begin{align}
&\Psicm =\Psicm *\Psicm ,
\label{eq:EOMm}
\\
&\cQ\Psicg +\Psicg *\Psicg=0 .
\label{eq:EOMg}
\end{align}

Here we shall fix our convention for the Neumann coefficient matrices.
The matter part of the three-string vertex defining the $*$-product is
given in the oscillator representation by
\begin{equation}
\ket{V^{\mathrm{m}}}_{123}=
\exp\!\Le(-\sum_{r,s=1}^3\sum_{m,n\geq 0}\half 
a_m^{(r)\dagger}V_{mn}^{rs}a_n^{(r)\dagger}
\Ri)\ket{p_1}\ket{p_2}\ket{p_3} ,
\label{eq:V}
\end{equation}
with $a_0=\sqrt{2}\,p$ (we are taking the convention of $\alpha'=1$).
The Neumann coefficient matrices and the vectors, $M_\alpha$ and
$\ve{v}_\alpha$ ($\alpha=0,\pm$), are related to $V^{rs}$ in
(\ref{eq:V}) as follows:
\begin{align}
&(M_0)_{mn}= (C V^{rr})_{mn},\quad
(M_\pm)_{mn}=(C V^{r,r\pm 1})_{mn} ,
\nn\\
&(\ve{v}_0)_n = V^{rr}_{n0},\quad
(\ve{v}_\pm)_n = V^{r,r\pm 1}_{n0} ,
\quad (m,n\ge 1)
\end{align}
where $C$ is the twist matrix, $C_{mn}=(-1)^m\delta_{mn}$.

The matter part solution $\Psicm$ to (\ref{eq:EOMm}) has been obtained
as a squeezed state \cite{KP:0008252,RSZ:0102112}:
\begin{equation}
\ket{\Psicm}=\left[\det(1-T\cM)\right]^{13}
\exp\biggl(-\half\sum_{m,n\ge 1}a_m^\dagger (CT)_{mn}a_n^\dagger
\biggr)\ket{0} ,
\end{equation}
where the matrix $T$ is given in terms of $M_0$ by
\begin{equation}
T=\frac{1}{2 M_0}\left(1+M_0-\sqrt{(1-M_0)(1+3M_0)}\right) .
\label{eq:T}
\end{equation}
The ghost part solution $\Psicg$ to (\ref{eq:EOMg}) has also
been obtained by taking the Siegel gauge and assuming the squeezed
state form \cite{HK:0108150}.
Beside determining $\Psicg$, (\ref{eq:EOMg}) fixes the
coefficients $f_n$ in $\cQ$ (\ref{eq:cQ}) which are arbitrary for the
gauge invariance alone.

Let us express the VSFT field $\Psi$ as a sum of $\Psic$ and the
fluctuation $\Phi$:
\begin{equation}
\Psi=\Psic + \Phi .
\label{eq:Psi=Psic+Phi}
\end{equation}
Then the linear part of the equation of motion for $\Phi$ reads
\begin{equation}
\wtcQ\Phi\equiv
\cQ\Phi+\Psic *\Phi+\Phi*\Psic=0 ,
\label{eq:wtcQ}
\end{equation}
where $\wtcQ$ is the BRST operator around the classical
solution $\Psic$.
We would like to construct the fluctuation modes $\Phi$ corresponding
to higher level open string states and satisfying the wave equation
(\ref{eq:wtcQ}).
We assume the factorization for these modes and that the ghost part is
common to that of $\Psic$:
\begin{equation}
\Phi=\Phim\otimes\Psicg .
\label{eq:Phi=PhimPsicg}
\end{equation}
Then the wave equation for the matter part $\Phim$ is given by
\begin{equation}
\Phim=\Psicm *\Phim +\Phim *\Psicm .
\label{eq:EOMf}
\end{equation}
In the following we are interested only in the matter part $\Phim$ of
the fluctuation modes and omit its superscript $\mathrm{m}$.
Eq.\ (\ref{eq:EOMf}) has been solved for the tachyon mode $\Phit$
\cite{HK:0108150}.
Explicitly, it is given by
\begin{equation}
\ket{\Phit}=\exp\biggl(
-\sum_{n\ge 1}t_n a_n^\dagger a_0 +ip\,\widehat x\biggr)
\ket{\Psicm} ,
\label{eq:tachyon}
\end{equation}
with
\begin{equation}
\ve{t}=3(1+T)(1+3M_0)^{-1}\ve{v}_0.
\end{equation}
It has been shown that $\Phit$ (\ref{eq:tachyon}) satisfies
(\ref{eq:EOMf}) when the momentum $p_\mu$ carried by $\Phit$ is on the
tachyon mass-shell $p_\mu^2=-m_{\rm tachyon}^2=1$.

Now we shall start constructing fluctuation modes at a generic mass
level.
As a candidate fluctuation mode with mass squared equal to $k-1$, let
us take the following one; $k$ creation operators acting on the
tachyon mode $\Phit$:
\begin{equation}
\ket{\Phik}=\sum_{n_1,\cdots,n_k\ge 1}
\beta^{\mu_1\cdots\mu_k}_{{n_1}\cdots{n_k}}
{a_{n_1}^{\mu_1\dagger}}\cdots{a_{n_k}^{\mu_k\dagger}}
\ket{\Phit} ,
\label{eq:massive-state}
\end{equation}
where $\beta$ is an unknown coefficient satisfying
\begin{equation}
\beta^*_{n_1\cdots n_k}=(-1)^k(-1)^{\sum_{i=1}^k n_i}
\beta_{n_1\cdots n_k} ,
\label{eq:bb*}
\end{equation}
which is due to the hermiticity constraint of $\Phik$
(see appendix \ref{evaluation}).
Substituting (\ref{eq:massive-state}) into (\ref{eq:EOMf}), we obtain
equations determining $\beta$. The detailed calculation using the
oscillator expression of the three-string vertex is presented in
appendix \ref{evaluation}.
Although the assumed state (\ref{eq:massive-state}) has fixed number $k$
of creation operators $a^\dagger$ acting on $\Phit$, there emerge on
the RHS of (\ref{eq:EOMf}) states with fewer number of
$a^\dagger$ acting on $\Phit$ besides those with $k$ $a^\dagger$s.
In order to eliminate these unwanted terms, we impose the following
transverse and traceless conditions on $\beta$:\footnote{
Due to these two conditions, (\ref{eq:transverse-condition}) and
(\ref{eq:traceless-condition}), our construction of massive modes is
restricted only to the highest spin states at a given mass level.
}
\begin{align}
&p_{\mu_1}\beta^{\mu_1\cdots\mu_k}=0,
\label{eq:transverse-condition}
\\
&\beta_{\mu_1}{}^{\mu_1\cdots\mu_{k-1}}=0 .
\label{eq:traceless-condition}
\end{align}
Then the equation for the coefficient $\beta$ is given by
\begin{equation}
\beta_{m_1\cdots m_k}- 2^{-p^2}
\Bigl(\rhom_{m_1n_1}\cdots \rhom_{m_kn_k}
+\rhop_{m_1n_1}\cdots \rhop_{m_kn_k}
\Bigr)\beta_{n_1\cdots n_k}
=0 ,
\label{eq:mass-eq}
\end{equation}
with
\begin{equation}
\rho_\pm=\frac{TM_\pm+M_\mp}{(1+T)(1-M_0)}.
\label{eq:rhopmcomm}
\end{equation}
The matrices $\rho_\pm$ are projection operators
\cite{Rastelli:2001rj} satisfying
\begin{equation}
\left(\rho_\pm\right)^2=\rho_\pm,\quad
\rho_+\rho_-=\rho_-\rho_+=0,\quad
\rho_+ +\rho_-=1 .
\label{eq:proj}
\end{equation}

Eqs.\ (\ref{eq:mass-eq}) and (\ref{eq:proj}) lead to a disappointing
result that our states (\ref{eq:massive-state}) can represent only
massless states. Namely, multiplying (\ref{eq:mass-eq}) by
$\rho_{s_1}\!\otimes\!\rho_{s_2}\!\otimes\cdots\otimes\!\rho_{s_k}$
with $s_i=+$ or $-$, we find that the equations for the purely
$\rho_+$  component $(\rho_+\otimes\cdots\otimes\rho_+)\beta$
and the purely $\rho_-$ one $(\rho_-\otimes\cdots\otimes\rho_-)\beta$ 
are reduced to
\begin{equation}
\bigl(1- 2^{-p^2}\bigr)(\rho_\pm\otimes\cdots\otimes\rho_\pm)
\beta=0 ,
\end{equation}
implying that they are massless states. On the other hand,
(\ref{eq:mass-eq}) tells that the mixed components, for example,
$(\rho_+\otimes\rho_-\otimes\cdots\otimes\rho_-)\beta$, are equal to
zero.
In the case of vector state with $k=1$, eq.\ (\ref{eq:mass-eq}) 
with $\rho_+ +\rho_-=1$ substituted reproduces the result of
\cite{HK:0108150} that this is a massless state and the coefficient
$\beta_n^\mu$ is arbitrary.

The above result is inevitable so long as the basic equations 
(\ref{eq:proj}) are valid. However, there is a subtle point concerning 
the eigenvalues of $\rho_\pm$.
Recall that the eigenvalue problem of the Neumann coefficient matrices
$M_0$ and $M_1\equiv M_+ - M_-$ has been solved in \cite{RSZ:0111281}.
They found that these matrices are expressed in terms of a single
matrix $K_1$ which is the matrix representation of the Virasoro
algebra $L_1+L_{-1}$, and the eigenvalue problem of $M_\alpha$ is
reduced to that of $K_1$.
Let $\ve{f}^{(\kappa)}$ be the eigenvector of $K_1$ corresponding to
the eigenvalue $\kappa$:
\begin{equation}
K_1\ve{f}^{(\kappa)}=\kappa\ve{f}^{(\kappa)} .
\label{eq:Kf=kf}
\end{equation}
The distribution of $\kappa$ is uniform and
extending from $-\infty$ to $\infty$.
This eigenvector $\ve{f}^{(\kappa)}$ is at the same time that of
$M_0$, $M_1$, $T$ and hence $\rho_\pm$. In particular we have
\begin{equation}
\rho_\pm\ve{f}^{(\kappa)}=\theta(\pm\kappa)\ve{f}^{(\kappa)} ,
\label{eq:rhof=kappaf}
\end{equation}
where $\theta(\kappa)$ is the step function
\begin{equation}
\theta(\kappa)=
\begin{cases}
1 & (\kappa > 0) \\ 0 & (\kappa <0)
\end{cases} .
\label{eq:theta}
\end{equation}
The subtle point is the eigenvalue of $\rho_\pm$ at $\kappa=0$.
In fact, it has been known that there is an eigenvector
$\ve{f}^{(\kappa=0)}$, which is twist-odd,
$C\ve{f}^{(0)}=-\ve{f}^{(0)}$.
However, the eigenvalue $\theta(\kappa=0)$ of $\rho_\pm$ is
indefinite. 

If we are allowed to set $\theta(0)=1/2$ in (\ref{eq:rhof=kappaf}),
which would look most plausible, the fluctuation (\ref{eq:massive-state})
represents a massive state at the expected mass level,
\begin{equation}
p^2=1-k ,
\end{equation}
by adopting either of the following two choices of
$\beta_{n_1\cdots n_k}^{\mu_1\cdots\mu_k}$ concerning
its dependence on the level number indices $n_1\cdots n_k$:
\begin{itemize}
\item $\beta$ is the tensor product of $k$ $\fz$s,
\begin{equation}
\beta=\fz\otimes\cdots\otimes\fz .
\label{eq:b=f...f}
\end{equation}
\item $\beta$ is the tensor product of $(k-1)$ $\fz$s and one
arbitrary vector $\ve{w}$,
\begin{equation}
\beta=\fz\otimes\cdots\otimes\fz\otimes\ve{w} .
\label{eq:b=f...fw}
\end{equation}
\end{itemize}
In both cases we have to multiply (\ref{eq:b=f...f}) and
(\ref{eq:b=f...fw}) by a transverse and traceless tensor carrying the
Lorentz indices, and carry out symmetrization if necessary.
Quite similarly, by taking $\beta$ which is a tensor product of $\ell$
$\fz$s and $k-\ell$ arbitrary vectors ($\ell \le k-1$), we obtain a
state at mass level $p^2=-\ell$.\footnote{
See sec.\ 4 for precise form of these states in the case $k=2$.
}

Now we have to resolve two problems. One is whether $\rho_\pm$ really
has eigenvalue $1/2$. Second, even if this is the case, we have
infinite degeneracy of massive as well as massless states which are
apparently physical ones.
We have to show that these spurious states are gauge artifacts.
Analysis of these two questions is the subject of the following two
sections.

\section{Numerical analysis of $\ve{\rho_\pm}$}

As seen in the previous section, the existence of the eigenvalue $1/2$ 
of the ``projection operators'' $\rho_\pm$ was essential for the
construction of the massive fluctuation modes.
The corresponding eigenvector is expected to be $\fz$, the zero-mode
of $K_1$.
It is obvious that we need some regularization for studying this
expectation since what we want to know is the value of the step
function $\theta(\kappa)$ (\ref{eq:theta}) at $\kappa=0$.
In the following, we shall show numerically that $\rho_\pm$ has indeed
eigenvalue $1/2$ by regularizing them to finite size matrices.

We have solved numerically the eigenvalue problem of the regularized
$\rho_+$ obtained by replacing $M_0$ and $M_1$ in it with $L\times L$
ones.
Since we have $\rho_-=C\rho_+ C$, we do not need to repeat the
analysis for $\rho_-$.
The expression (\ref{eq:rhopmcomm}) of $\rho_\pm$ was in fact
obtained by naively using the non-linear relations among the Neumann
coefficient matrices \cite{GJ1,GJ2} upon the original expressions,
which are given by (see appendix \ref{evaluation})
\begin{equation}
\rho_\pm =(M_+,M_-)(1-T\cM)^{-1}
\begin{cases}
\sPmatrix{0\\ 1}\\\sPmatrix{1\\ 0}
\end{cases} ,
\label{eq:rhopmorig}
\end{equation}
with
\begin{equation}
\cM=\begin{pmatrix}
M_0 & M_+ \\ M_- & M_0
\end{pmatrix} .
\label{eq:cM}
\end{equation}
Since the non-linear relations no longer hold for regularized
$M_\alpha$ and naive use of them may be dangerous near $\kappa=0$
\cite{Anomaly,HatMorTer,Reexam}, we have employed the original
expression (\ref{eq:rhopmorig}) in our numerical analysis.

Tables \ref{rho+-even-eigenvalue} -- \ref{rho+-odd-eigenvector}
show the result of our calculations.
Since the eigenvalue distributions are qualitatively different between
even and odd $L$, we have carried out the analysis for each of these
two cases.
In the case of even $L$, all the eigenvalues of $\rho_+$ are close to
either $0$ or $1$ except two ``anomalous'' ones, $\lambda^{(1)}$ and
$\lambda^{(2)}$ , which are given
in table \ref{rho+-even-eigenvalue} for various even $L$.
Though the raw values of these anomalous eigenvalues are not so close
to $1/2$, their values at $L=\infty$ obtained by fitting are
surprisingly close to the expected value of $1/2$.

\begin{table}[htbp]
\begin{center}
\begin{tabular}{|c||c|c|}
\hline
 $L$  & $\lambda^{(1)}$ & $\lambda^{(2)}$\\\hline\hline
 $50$  & $0.771$ & $0.259$\\\hline
 $100$ & $0.752$ & $0.279$\\\hline
 $150$ & $0.742$ & $0.289$\\\hline
 $200$ & $0.735$ & $0.295$\\\hline
 $300$ & $0.726$ & $0.304$\\\hline
 $500$ & $0.716$ & $0.314$\\\hline
$\infty$ &$0.512$& $0.489$\\\hline
\end{tabular}
\end{center}
\caption{Anomalous eigenvalues of $\rho_+$ (\ref{eq:rhopmorig})
for various even $L$. The values at $L=\infty$ have been obtained 
by the fitting function of the form
$\sum_{k=0}^5 c_k/\left(\ln L\right)^k$. We use the same fitting
function also in other tables \ref{rho+-even-eigenvector1}
-- \ref{rho+-odd-eigenvector}.
}
\label{rho+-even-eigenvalue}
\end{table}

\begin{table}[htbp]
\begin{center}
\begin{tabular}{|c||c|c|c|c|c||c|c|c|c|c|}
\hline
 $L$ & $a^{(1)}_3$ & $a^{(1)}_5$ & $a^{(1)}_7$ & $a^{(1)}_9$ &
 $a^{(1)}_{11}$ & 
 $b^{(1)}_4$ & $b^{(1)}_6$ & $b^{(1)}_8$ & $b^{(1)}_{10}$ &
 $b^{(1)}_{12}$ \\\hline\hline
 $50$  & $1.104$ & $1.185$ & $1.256$ & $1.320$ & $1.382$ &
 $-1.073$ & $1.119$ & $-1.030$ & $1.208$ & $-1.257$\\\hline
 $100$ & $1.086$ & $1.150$ & $1.204$ & $1.251$ & $1.295$&
 $-1.040$ & $1.055$ & $-1.067$ & $1.079$ & $-1.092$\\\hline
 $150$ & $1.077$ & $1.134$ & $1.183$ & $1.223$ & $1.260$&
 $-1.026$ & $1.029$ & $-1.030$ & $1.032$ & $-1.035$\\\hline
 $200$ & $1.072$ & $1.125$ & $1.168$& $1.205$ & $1.239$&
 $-1.018$ & $1.015$ & $-1.010$ & $1.007$ & $-1.005$\\\hline
 $300$ & $1.065$ & $1.113$ & $1.151$& $1.185$ & $1.215$&
 $-1.009$ & $0.998$ & $-0.987$ & $0.977$ & $-0.970$\\\hline
 $500$ & $1.058$ & $1.100$ & $1.134$ & $1.163$ & $1.189$&
 $-1.000$ & $0.981$ & $-0.964$ & $0.949$ & $-0.937$\\\hline
 $\infty$&$0.999$&$1.003$& $1.008$ & $1.012$ & $1.008$&
 $-0.944$ & $0.883$ & $-0.819$ & $0.735$ & $-0.606$\\\hline
\end{tabular}
\end{center}
\caption{
Components of the vectors $\ve{a}^{(1)}$ and $\ve{b}^{(1)}$ for
various even $L$. 
}
\label{rho+-even-eigenvector1}
\end{table}

\begin{table}[htbp]
\begin{center}
\begin{tabular}{|c||c|c|c|c|c||c|c|c|c|c|}
\hline
 $L$ & $a^{(2)}_3$ & $a^{(2)}_5$ & $a^{(2)}_7$ & $a^{(2)}_9$ &
 $a^{(2)}_{11}$ & 
$b^{(2)}_4$ & $b^{(2)}_6$ & $b^{(2)}_8$ & $b^{(2)}_{10}$ &
 $b^{(2)}_{12}$ \\\hline\hline
 $50$  & $1.064$ & $1.114$ & $1.159$ & $1.201$ & $1.243$ &
$-1.030$ & $1.038$ & $-1.047$ & $1.058$ & $-1.074$ \\\hline
 $100$ & $1.056$& $1.096$ & $1.130$ & $1.161$ & $1.189$ &
$-1.014$ & $1.006$ & $-0.998$ & $0.992$ & $-0.988$ \\\hline
 $150$ & $1.052$ & $1.089$ & $1.112$ & $1.146$ & $1.170$ &
$-1.007$ & $0.994$ & $-0.981$ & $0.970$ & $-0.961$ \\\hline
 $200$ & $1.049$ & $1.084$ & $1.113$ & $1.138$ & $1.160$ &
$-1.003$ & $0.987$ & $-0.972$ & $0.958$ & $-0.947$ \\\hline
 $300$ & $1.046$ & $1.079$ & $1.105$ & $1.128$ & $1.147$ &
$-0.999$ & $0.979$ & $-0.960$ & $0.944$ & $-0.930$ \\\hline
 $500$ & $1.043$ & $1.072$ & $1.096$ & $1.117$ & $1.134$&
$-0.994$ & $0.970$ & $-0.949$ & $0.930$ & $-0.913$ \\\hline
 $\infty$&$0.991$&$0.988$& $0.987$ & $0.983$ & $0.970$&
$-0.941$ & $0.888$ & $-0.840$ & $0.785$ & $-0.704$ \\\hline
\end{tabular}
\caption{
Components of the vectors $\ve{a}^{(2)}$ and $\ve{b}^{(2)}$ for
various even $L$. 
}
\label{rho+-even-eigenvector2}
\end{center}
\end{table}

Analysis of the eigenvectors of these anomalous eigenvalues is
presented in tables \ref{rho+-even-eigenvector1} and
\ref{rho+-even-eigenvector2}.
Let us denote by $\ve{u}^{(1)}$ and $\ve{u}^{(2)}$ the eigenvectors
corresponding to the eigenvalues $\lambda^{(1)}$ and $\lambda^{(2)}$,
respectively. These eigenvectors do not have a definite twist.
We define a twist-odd vector $\ve{a}^{(i)}$ ($i=1,2$) with components
$a_{2n+1}^{(i)}=u_{2n+1}^{(i)}/f^{(0)}_{2n+1}$,
and a twist-even one $\ve{b}^{(i)}$ with
$b_{2n}^{(i)}=u_{2n}^{(i)}/u_{2}^{(i)}$.
Here, $\ve{u}^{(i)}$ is normalized so that $u_1^{(i)}=1$, and the
components of $\fz$ are given by \cite{RSZ:0111281}
\begin{equation}
f^{(0)}_n =
\begin{cases}
\displaystyle
\frac{(-1)^{(n-1)/2}}{\sqrt{n}} & n\mbox{: odd} \\
0 & n\mbox{: even}
\end{cases}
\label{eq:fz_n}
\end{equation}
Components of $\ve{a}^{(i)}$ and $\ve{b}^{(i)}$ for various even $L$
and their values at $L=\infty$ obtained by fitting are given in tables 
\ref{rho+-even-eigenvector1} and \ref{rho+-even-eigenvector2} for
$i=1$ and $2$, respectively.
They strongly support that all (odd) components of $\ve{a}^{(i)}$ are
equal to one and that $\ve{b}^{(1)}$ and $\ve{b}^{(2)}$ are equal to
each other.\footnote{
If we adopt the fitting by polynomials of $1/L$, we obtain better
coincidence between $b_{2n}^{(1)}$ and $b_{2n}^{(2)}$ at $L=\infty$
for larger $n$.
}
This implies that $\ve{u}^{(1)}$ and $\ve{u}^{(2)}$ are given as
linear combinations of two vectors, $\fz$ which is twist-odd and
$\ve{b}^{(1)}(=\ve{b}^{(2)})$ which is twist-even.
Therefore, in the limit $L\to\infty$, the projection operator $\rho_+$ 
has the eigenvalue $1/2$ which is doubly degenerate, and the
corresponding eigenvectors are $\fz$ and $\ve{b}^{(1)}$.
The norm of $\ve{b}^{(1)}$ seems to have worse divergence than that of 
$\fz$ which is logarithmically divergent.
We do not know whether $\ve{b}^{(1)}$, which appears only for even
$L$, has any relevance to the construction of fluctuation modes.

\begin{table}[htbp]
\begin{center}
\begin{tabular}{|c||c|}
\hline
 $L$  & $\lambda$\\\hline\hline
 $49$   & $0.525$ \\\hline
 $99$   & $0.523$ \\\hline
 $149$  & $0.522$ \\\hline
 $199$  & $0.521$ \\\hline
 $299$  & $0.520$ \\\hline
 $499$  & $0.519$ \\\hline
$\infty$&$0.501$\\\hline
\end{tabular}
\end{center}
\caption{Anomalous eigenvalues of $\rho_+$ (\ref{eq:rhopmorig})
for various odd $L$ and their extrapolation to $L=\infty$.
}
\label{rho+-odd-eigenvalue}
\end{table}

\begin{table}[htbp]
\begin{center}
\begin{tabular}{|c||c|c|c|c|c|}
\hline
 $L$ & $a_3$ & $a_5$ & $a_7$ & $a_9$  & $a_{11}$ \\\hline\hline
 $49$  & $1.104$ & $1.187$ & $1.262$ & $1.332$ & $1.402$\\\hline
 $99$  & $1.084$ & $1.147$ & $1.201$ & $1.249$ & $1.293$\\\hline
 $149$ & $1.075$ & $1.131$ & $1.178$ & $1.217$ & $1.255$\\\hline
 $199$ & $1.069$ & $1.121$ & $1.163$ & $1.200$ & $1.232$\\\hline
 $299$ & $1.063$ & $1.109$ & $1.146$ & $1.179$ & $1.208$\\\hline
 $499$ & $1.056$ & $1.097$ & $1.123$ & $1.158$ & $1.182$\\\hline
$\infty$&$0.997$ & $1.000$ & $1.001$ & $0.993$ & $0.961$\\\hline
\end{tabular}
\caption{Components of the vector $\ve{a}$ for various odd $L$ and 
 their extrapolation to $L=\infty$.
}
\label{rho+-odd-eigenvector}
\end{center}
\end{table}

In the case of odd $L$, we have only one anomalous eigenvalue
$\lambda$ largely deviated from either $0$ or $1$. Table
\ref{rho+-odd-eigenvalue} shows this eigenvalue $\lambda$ for various
odd $L$ and its extrapolation to $L=\infty$.
As expected, the table \ref{rho+-odd-eigenvalue} supports that
$\lambda\to 1/2$ as $L\to\infty$.
The eigenvector $\ve{u}$ of this eigenvalue $\lambda$ does not have
definite a twist for a finite $L$. However, its twist-even component
is negligibly small compared with the twist-odd one.
In fact, the norm of the twist-even part, $(1+C)\ve{u}/2$, is at most
$0.6\%$ of the norm of the whole vector $\ve{u}$.\footnote{
In the case of even $L$, the norms of the even and odd parts of
$\ve{u}^{(i)}$, $(1\pm C)\ve{u}^{(i)}/2$, are of
the same order.
}
Therefore, we have studied only the twist-odd vector $\ve{a}$ with
components $a_{2n+1}=u_{2n+1}/f^{(0)}_{2n+1}$.
The results given in table \ref{rho+-odd-eigenvector} again support
our expectation that $\ve{u}$ is equal to $\fz$.

In summary, in both even and odd $L$ cases, our numerical analysis of
the eigenvalue problem of $\rho_\pm$ (\ref{eq:rhopmorig}) confirms our 
expectation that $\rho_\pm$ has an anomalous eigenvalue $1/2$ and the
corresponding eigenvector is $\fz$.

\section{Gauging away the spurious states}

In sec.\ \ref{section-massive} we have shown that we can construct
massive fluctuation modes if the projection operator $\rho_\pm$ has
eigenvalue $1/2$. This property has been verified numerically in
the last section.
However, there still remains a problem: we have infinite degeneracy of 
massive and massless states as we saw in the last part of sec.\
\ref{section-massive}.
In this section we shall show that these spurious states can in fact
be gauged away.
Our argument here is an application of that given in
\cite{Imamura:0204031} for spurious massless vector states.

For simplicity we restrict ourselves to the lowest massive state
$\Phi^{(k=2)}$:
\begin{equation}
\ket{\Phi^{(2)}}=\beta_{mn}^{\mu\nu}
a^{\mu\dagger}_m a^{\nu\dagger}_n \ket{\Phit} ,
\label{eq:Phi2}
\end{equation}
where $\beta_{mn}^{\mu\nu}$ is traceless and transverse with respect
to its Lorentz indices.
However, the following argument can straightforwardly be extended to
$\Phik$ with larger $k$.
Fourier-expanding $\beta_{mn}$ in terms of the eigenvector
$\ve{f}^{(\kappa)}$ of $K_1$,
\begin{equation}
\beta_{mn}^{\mu\nu}=\int\!d\kappa\int\!d\lambda\,
f^{(\kappa)}_m f^{(\lambda)}_n \bkl,
\label{eq:intrepbeta}
\end{equation}
and substituting it into the equation for $\beta$, (\ref{eq:mass-eq})
with $k=2$, we obtain the following equation for
$\bkl$:
\begin{equation}
\left(1 - 2^{-p^2}\left[\theta(-\kappa)\theta(-\lambda)
+\theta(\kappa)\theta(\lambda)\right]\right)\bkl=0 .
\label{eq:eombetakl}
\end{equation}
The construction of massive states given in sec.\
\ref{section-massive} (see the paragraph containing
(\ref{eq:b=f...f}) and (\ref{eq:b=f...fw})) can be restated as follows
in terms of the spectral function $\bkl$.
First, (\ref{eq:eombetakl}) implies that $\bkl$ must
have support only in the regions $\{\kappa\ge 0,\lambda\ge 0\}$
and $\{\kappa\le 0,\lambda\le 0\}$.
Taking into account that $\theta(0)=1/2$ as we have seen in the
previous section, there are three possible types of $\bkl$ giving
massive as well as massless states (table \ref{tab:bkl}).
Since the function $g(\kappa)$ for the type-B state and $\bkl$ for the 
type-C are arbitrary except that they are smooth (and has support
only in the regions stated above for $\bkl$ for type-C), these two
types of states are infinitely degenerate.
We shall show that the type-B and C states can in fact be removed by
gauge transformations of VSFT.

\renewcommand{\arraystretch}{1.3}
\begin{table}[htbp]
\begin{center}
\begin{tabular}[b]{|c|c|c|}
\hline
&$\bkl$ & $\mbox{(mass)}^2$\\
\hline\hline
A &$\delta(\kappa)\delta(\lambda)$ & $1$ \\
\hline
B &$\delta(\kappa)g(\lambda)$ & $1$ \\
\hline
C &arbitrary smooth function & 0 \\
\hline
\end{tabular}
\caption{Three types of $(\kappa,\lambda)$-dependence of $\bkl$
  satisfying (\ref{eq:eombetakl})
and the corresponding mass squared. The function $g(\kappa)$ is an
arbitrary smooth function.
The whole $\bkl$ must be symmetric under the exchange of
$(\mu,\kappa)$ and $(\nu,\lambda)$.
}
\label{tab:bkl}
\end{center}
\end{table}
\renewcommand{\arraystretch}{1}

The gauge transformation of VSFT, (\ref{eq:dLPsi}), expressed in terms 
of the fluctuation $\Phi$ (\ref{eq:Psi=Psic+Phi}) reads
$\delta_\Lambda\Phi=\wtcQ\Lambda+\Phi *\Lambda-\Lambda *\Phi$
with $\wtcQ$ defined by (\ref{eq:wtcQ}).
We shall consider the inhomogeneous part
\begin{equation}
\delta_\Lambda^{\mathrm{I}}\Phi=\wtcQ\Lambda ,
\label{eq:dLI}
\end{equation}
of the whole transformation, and in particular take the following type
of $\Lambda$:
\begin{equation}
\Lambda= \Lambda^{\mathrm{m}}\otimes \cI^{\mathrm{g}}, 
\label{eq:trf-form}
\end{equation}
where $\cI^{\mathrm{g}}$ is the ghost part of the identity string
field $\cI=\cI^{\mathrm{m}}\otimes\cI^{\mathrm{g}}$
satisfying $\cI^{\mathrm{g}}*\Psicg=\Psicg*\cI^{\mathrm{g}}=\Psicg$
for $\Psicg$ of (\ref{eq:EOMg})
and $\cQ\,\cI^{\mathrm{g}}=0$
\cite{Kishimoto:0110124,KO:0112169,Imamura:0204031}.
For this $\Lambda$ and fluctuation $\Phi$ of the factorized form
(\ref{eq:Phi=PhimPsicg}), we have
\begin{equation}
\delta_\Lambda^{\mathrm{I}}\Phim
= \Psicm *\Lambda^{\mathrm{m}}-\Lambda^{\mathrm{m}}*\Psicm .
\label{eq:dLIPhim}
\end{equation}
As the matter part $\Lambda^{\mathrm{m}}$ of the gauge transformation
string field, we take
\begin{equation}
\ket{\Lambda^{\mathrm{m}}}
= \gamma_{mn}^{\mu\nu} a^{\mu\dagger}_m a^{\nu\dagger}_n
\ket{\Phit} ,
\label{eq:L=gaaPhit}
\end{equation}
with the coefficient $\gamma_{mn}^{\mu\nu}$ being traceless and
transverse with respect to $\mu$ and $\nu$.
Then, (\ref{eq:dLIPhim}) is given by
\begin{equation}
\delta_\Lambda^{\mathrm{I}}\ket{\Phim} = 2^{-p^2}
\Bigl(\rhom_{mp}\rhom_{nq}-\rhop_{mp}\rhop_{nq}\Bigr)
\gamma_{pq}^{\mu\nu}a^{\mu\dagger}_m a^{\nu\dagger}_n
\ket{\Phit} ,
\label{eq:dLIPhimgamma}
\end{equation}
which is expressed as the following transformation on the spectral
function $\bkl$ of (\ref{eq:intrepbeta}):
\begin{equation}
\delta_\Lambda^{\mathrm{I}}\bkl= 2^{-p^2}
\Bigl(\theta(-\kappa)\theta(-\lambda)-\theta(\kappa)\theta(\lambda)
\Bigr)\gkl ,
\label{eq:dLbeta}
\end{equation}
where $\gkl$ is defined for $\gamma_{mn}^{\mu\nu}$ similarly to
(\ref{eq:intrepbeta}).

Eq.\ (\ref{eq:dLbeta}) implies that the type-B and C states are
unphysical ones which can be eliminated by the present gauge
transformation. First, the type-B states are gauged away by taking
$\delta(\lambda)g(\kappa)\epsilon(\kappa)$ as the
$(\kappa,\lambda)$-dependence of $\gkl$. Here, $\epsilon(\kappa)$ is
the signature function
$\epsilon(\kappa)=\theta(\kappa)-\theta(-\kappa)$.
Second, $\gkl$ for gauging away the type-C states is
$\gkl=-2^{p^2}\left[\theta(-\kappa)\theta(-\lambda)-
\theta(\kappa)\theta(\lambda)\right]\bkl$.
It is obvious that the type-A states cannot be removed by the present
gauge transformation.

Finally in this section we shall comment on the relation between the
gauge transformation used in this paper and that in
\cite{Imamura:0204031}.
Our gauge transformation string field $\Lambda^{\mathrm{m}}$
(\ref{eq:L=gaaPhit}) is of different type from that used in
\cite{Imamura:0204031}; the latter is based on the identity string
field instead of the classical solution $\Psicm$.
If we have adopted $\Lambda^{\mathrm{m}}$ of the type of
\cite{Imamura:0204031}, we would have obtained (\ref{eq:dLIPhimgamma})
with $\rho_\pm$ replaced by $(1+T)\rho_\pm$.
This gauge transformation cannot remove the type-B states
which are absent for the vector case discussed in
\cite{Imamura:0204031}.

\section*{Acknowledgments}
We would like to thank K.\ Hashimoto, Y.\ Imamura, E.\ Itou,
H.\ Kajiura, I.\ Kishimoto, Y.\ Matsuo, S.\ Moriyama, T.\ Muramatsu,
K.\ Ohmori, T.\ Takahashi and S.\ Teraguchi for valuable discussions
and comments.
The works of H.\,H. was supported in part by a
Grant-in-Aid for Scientific Research from Ministry of Education,
Culture, Sports, Science, and Technology (\#12640264).

\appendix

\section{Derivation of (\ref{eq:mass-eq})}
\label{evaluation}

In this appendix we outline a derivation of (\ref{eq:mass-eq}) from
the equation of motion (\ref{eq:EOMf}).
First we shall mention the hermiticity constraint (\ref{eq:bb*}).
We impose the following hermiticity condition on the matter part 
$\Phim$ of a string field of the type (\ref{eq:Phi=PhimPsicg}):
\begin{equation}
{}_2\bra{\Phim}=\prod_{r=1,2}\int\!\frac{d^{26}p_r}{(2\pi)^{26}}
\,(2\pi)^{26}\delta^{26}(p_1+p_2)
\,{}_{12}\bracket{R^{\mathrm{m}}}{\Phim}_1 ,
\label{eq:ref-beta-k}
\end{equation}
where the matter reflector (two-string vertex) $\bra{R^{\mathrm{m}}}$
is defined by
\begin{equation}
{}_{12}\bra{R^{\mathrm{m}}}=\bra{p_1}\bra{p_2}
\exp\biggl(-\sum_{n\geq 1}(-1)^n a^{(1)}_na^{(2)}_n\biggr) .
\label{eq:R}
\end{equation}
This constraint reduces the number of degrees of freedom in $\Phim$ to
half and ensures the hermiticity of the action.
Eq.\ (\ref{eq:bb*}) for $\beta$ is immediately obtained by plugging
the expression (\ref{eq:massive-state}) into (\ref{eq:ref-beta-k})
and using that $\Phit$ itself satisfies (\ref{eq:ref-beta-k}).

The wave equation (\ref{eq:EOMf}) for the matter
fluctuation $\Phim$ is rewritten in the oscillator representation as
\begin{equation}
\ket{\Phim}_3-{}_1\bra{\Psicm}_2\bracket{\Phim}{\Vm}_{123}
-{}_1\bra{\Phim}_2\bracket{\Psicm}{\Vm}_{123}=0 ,
\label{eq:EOMV}
\end{equation}
where we have omitted the integrations
$\left(\prod_{r=1}^3\int d^{26}p_r/(2\pi)^{26}\right)
(2\pi)^{26}\delta^{26}(p_1+p_2+p_3)$
in the second and third terms.
Let us consider the second term of (\ref{eq:EOMV}) with $\Phik$
(\ref{eq:massive-state}) substituted for $\Phim$,
${}_1\bra{\Psicm}_2\bracket{\Phik}{\Vm}_{123}$.
The basic formula for this calculation is obtained from the well-known 
formula valid for any bosonic oscillators satisfying
$[a_i,a^\dagger_j]=\delta_{ij}$:
\begin{align}
& \lvac \exp \Le (  - \half a_i A_{ij} a_{j}-
K_ia_i \Ri )
  \exp \Le ( -\half a_i^\dagger B_{ij}
  a_{j}^\dagger -J_ia_i^\dagger\Ri )\rvac\nn\\
&\qquad=[\det \Le (1-AB\Ri )]^{-1/2}
\exp \Le (-\half JPAJ-\half KBPK+JPK \Ri ) ,
\label{eq:exp}
\end{align}
with $P=(1-AB)^{-1}$.
Letting $(-\del/\del K_{i_1})\cdots(-\del/\del K_{i_k})$ act on
(\ref{eq:exp}), we get
\begin{align}
&\lvac a_{i_1}\cdots a_{i_k}
\exp\!\left(-\half a_i A_{ij} a_{j}-K_ia_i\right)
\exp\!\left(-\half a_i^\dagger B_{ij}
  a_{j}^\dagger -J_ia_i^\dagger\right)\rvac\nn\\
&=[\det \Le (1-AB\Ri)]^{-\half}\Biggl\{
\prod_{a=1}^k(KBP-JP)_{i_a}
\nn\\
&\qquad
+\sum_{a>b}(BP)_{i_a i_b}\prod_{c\ne a,b}(KBP-JP)_{i_c}
+ \ldots \Biggr\}
\exp \Le (-\half JPAJ-\half KBPK+JPK \Ri ) ,
\label{eq:k-differentiation}
\end{align}
where we have omitted terms with more than one $(BP)_{i_a i_b}$
factors.
In our applications of (\ref{eq:k-differentiation}), the index $i$
represents the level number $n$, the Lorentz index $\mu$ and the
string index $r=1,2$.
Contracting (\ref{eq:k-differentiation}) with
$(\beta^{\mu_1\cdots\mu_k}_{n_1\cdots n_k})^*$,
the terms containing the $(BP)_{i_a i_b}$ factors drop out due to the 
traceless condition (\ref{eq:traceless-condition}), while due to the
transverse condition the $(KBP-JP)_{i}$ factor with $i=(n,\mu,r=2)$
contributes only $-a_m^{\mu (3)\dagger}\!\left(\rho_- C\right)_{mn}$
with $\rho_-$ defined by (\ref{eq:rhopmorig}).
Therefore, we have
\begin{align}
{}_1\bra{\Psicm}_2\bracket{\Phik}{\Vm}_{123}
&=2^{-p^2}
(-1)^k a^{\mu_1(3)\dagger}_{m_1}\cdots a^{\mu_k(3)\dagger}_{m_k}
(\rho_-C)_{m_1n_1}\cdots(\rho_-C)_{m_kn_k}
(\beta^{\mu_1\cdots\mu_k}_{n_1\cdots n_k})^*\ket{\Phit}
\nn\\
&=2^{-p^2}
a^{\mu_1(3)\dagger}_{m_1}\cdots a^{\mu_k(3)\dagger}_{m_k}
\rhom_{m_1n_1}\cdots \rhom_{m_kn_k}
\beta^{\mu_1\cdots\mu_k}_{n_1\cdots n_k}\ket{\Phit} ,
\end{align}
where we have used $\Psicm *\Phit=2^{-p^2}\Phit$ and the hermiticity
(\ref{eq:bb*}).
This gives the $\rho_-$ term of (\ref{eq:mass-eq}).
Derivation of the $\rho_+$ term of (\ref{eq:mass-eq}) from the last
term of (\ref{eq:EOMV}) is quite similar.

\end{document}